\def\ltsima{$\; \buildrel < \over \sim \;$}
\def\lsim{\lower.5ex\hbox{\ltsima}}
\def\gtsima{$\; \buildrel > \over \sim \;$}
\def\gsim{\lower.5ex\hbox{\gtsima}}
\def\hh{\hbox{$^h$}}
\def\mm{\hbox{$^m$}}
\def\ss{\hbox{$^s$}}
\def\degr{\hbox{$^\circ$}}
\def\arcmin{\hbox{$^\prime$}}
\def\arcsec{\hbox{$^{\prime\prime}$}}

\documentstyle{l-aa}

\begin{document}

   \thesaurus{
               (09.11.1;  
                19.50.1;  
                24.02.1)  
}
%
   \title{Search for Old Neutron Stars in molecular clouds:
	  Cygnus Rift and Cygnus OB7}

   \author{T. Belloni \inst{1}
   \and L. Zampieri \inst{2} \thanks{
Present address: University of Illinois at Urbana--Champaign, Department
of Physics, Loomis Laboratory of Physics, 1110 West Green Street, Urbana,
Illinois 61801-3080, U.S.A.
}
   \and S. Campana \inst{3}}

   \offprints{T. Belloni}

   \institute{\inst{1} Astronomical Institute ``Anton Pannekoek'',
	      University of Amsterdam and Center for High-Energy Astrophysics,
	      Kruislaan 403, NL-1098 SJ Amsterdam, The Netherlands\\
	      \inst{2} International School for Advances Studies (S.I.S.S.A.),
	      Via Beirut 2--4, I-34013 Trieste, Italy\\
	      \inst{3} Osservatorio Astronomico di Brera,
	      Via Bianchi 46, I-22055 Merate, Italy}

   \date{Received ... ; accepted ...}

   \maketitle

   \begin{abstract}
We present the results of a systematic search for
old isolated neutron stars (ONSs) in the direction of two
giant molecular clouds in Cygnus (Rift and OB7).
From theoretical calculations, we expect the detection of a
large number of ONSs with the PSPC on board ROSAT.
By analyzing the PSPC pointings in the direction of the clouds,
we find four sources characterized by count rates ($\sim
10^{-3}$ c s$^{-1}$) and spectral properties consistent with the hypothesis
that the X--ray radiation is produced by ONSs
and also characterized by the absence of any
measurable optical counterpart within their error circle in the
digitized red plates of the Palomar All Sky Survey. The importance of
follow--up deep observations in the direction of these ONS candidates is
discussed.
The observational and theoretical approach presented here could be
fruitfully applied also to the systematic search for ONSs in other
regions of the Galaxy.

 \end{abstract}

%

\section{Introduction}

In the first phases following their formation, neutron stars are probably
strongly magnetized and can emit intense electromagnetic dipole radiation
at the expense of their rotational energy. These energy losses will
produce a non--negligible radiation pressure on the surrounding medium,
inhibiting any possible accretion of interstellar material (see e.g.
Lipunov 1992).
After the radiation pressure has dropped, the flow penetrates
inside the accretion radius, proceeding unaffected until the
magnetospheric (Alfv\`en) radius is reached, where the magnetic pressure
becomes equal to the ram pressure of the infalling gas.
At this point accretion can continue if the centrifugal acceleration
exerted on the matter flowing along the field lines is smaller than the
gravitational acceleration: in order to have accretion from the
Interstellar Medium (ISM) a relic magnetic field $B \lsim 10^9$ G and a
rotational period greater than a few seconds are needed (Illarionov \&
Sunyaev 1975), which are not implausible for Old Neutron Stars (ONSs).
If this condition is satisfied, ONSs should appear as weak, soft
X--ray sources, as firstly suggested by Ostriker, Rees \& Silk (1970).

The detection of ONSs is a long-sought goal of X--ray astronomy such that
it was included as a possible target for the {\it Einstein\/} mission
(Helfand, Chanan \& Novick 1980), but the first systematic study of the
observability of these sources with ROSAT was carried out by Treves \&
Colpi (1991), who found that, in the most favourable case
of polar cap accretion, thousands of ONSs should appear in the ROSAT PSPC
All Sky Survey. A complete analysis by Blaes \& Madau (1993) essentially
confirmed the results of Treves \& Colpi (1991).
Zane et al. (1995a) reconsidered the detectability of
ONSs with ROSAT and 
found that, for polar cap accretion, ONSs may contribute up to 10\%
of the unresolved X--ray background, although in this case about 10
sources deg$^{-2}$ should be observable by ROSAT at the sensitivity limit
of $10^{-3}$ c s$^{-1}$. Finally, in two recent
investigations Shemi (1995) and Zane et al. (1995b) have studied
the detectability of nearby ONSs.
To date, good evidence for ONS candidates has been presented by Stocke
et al. (1995) and Walter, Wolk \& Neuh\"auser (1996), who found that
two unidentified X--ray sources detected in either the {\it
Einstein\/} Extended Medium Sensitivity Survey or in the ROSAT All
Sky Survey may actually be powered by an accreting ONS.

A different approach for the search of ONSs was suggested by Blaes \&
Madau (1993) and Colpi, Campana \& Treves (1993), who indicate as a very
promising environment for their detection nearby giant molecular clouds,
where the density of the ISM is sufficiently high for a single neutron
star to emit a considerable amount of radiation.
However, in this case also the absorption of the ISM will be enhanced and
the detectability will crucially rely on the delicate balance
between the emitted luminosity and the absorbed flux. As shown by
Blaes \& Madau (1993), Colpi,
Campana \& Treves (1993) and Zane et al. (1995a), some molecular clouds in
the vicinity of the Sun are expected to contain a large number of ONSs
emitting at relatively high rates and then representing the most
favourable sites for the observability of these sources.

On the wake of these studies, we have performed a
systematic investigation of the 
X--ray sources
detected by ROSAT PSPC in the direction of two molecular clouds (Cygnus
Rift and Cygnus OB7), which should be particularly favourable as far as
the expected number of ONSs is concerned. As we will see, the definition
of a systematic searching strategy is very important when dealing with
large number of sources and, to this end, the theoretical investigation
of the expected emission properties turns out to be certainly a key
ingredient.

\begin{table*}
\caption{Characteristic parameters for Cygnus Rift and Cygnus OB7}
\begin{tabular}{lccccc}

Cloud & volume $V_c$ & apparent surface $\Sigma_c$ & inner radius $r_{in}$
& outer radius $r_{out}$ & density $n_c$ \\
      & (pc$^3$) & (deg$^2$) & (pc) & (pc) & (cm$^{-3}$) \\
\hline
Cygnus Rift & 1.26 $\times 10^6$ & 112 & 662 & 738 & 29\\
Cygnus OB7  & 1.10 $\times 10^6$ &  46 & 739 & 861 & 29\\
\end{tabular}
\end{table*}

Section 2 is devoted to the theoretical analysis of the
number of ONSs detectable by ROSAT PSPC in the direction of these two
molecular clouds and their emission properties, whereas in
section 3 we will present the searching procedure
and the four ONS candidates which have been found in this way.
Finally, section 4 contains a discussion of these results
in connection with the detectability issue of ONSs and a suggestion for
possible new deep observations to shed further light on the nature of
these objects.

\section{Theoretical expectations}

\subsection{Assumptions}

For a star moving supersonically with velocity $v$ relative to the
interstellar gas, the amount of mass which is accreted per unit time is
${\dot M} \simeq \pi r_{gc}^2 n m_p v$,
where $r_{gc} = 2 \chi GM_*/v^2$ is the gravitational capture radius,
$M_*$ is the mass of the neutron star, $n$ the number density of the
ISM ($n = 1.36\, n_H$, where $n_H$ is the hydrogen number density, for
standard chemical composition) and $\chi$ is parameter of order
unity which encompasses the theoretical uncertainties on the value of
$r_{gc}$. In the following we take $\chi$ = 1 (see e.g. Novikov \& Thorne
1973).
At the typical values of the ISM density (outside molecular clouds),
it is not clear that the fluid approximation applies at the gravitational
capture radius (see Koester 1976) and
the accretion rate could be greatly reduced. However, low magnetic
fields, frozen in the interstellar medium, may effectively couple the
gas particles. In this case, if the Larmor radius is smaller than the
gravitational capture radius, the fluid approximation can be still
considered valid and the accretion rate is reasonably expressed
through the previous formula.
The total luminosity $L$ emitted by an ONS is then
\begin{eqnarray}
L & = & \eta \ {\dot M} \,c^2 = 9.5\times 10^{29}
\left( {\eta\over 0.18} \right) \nonumber \\
& & \left( {M_*\over {M_\odot}} \right)^2
\left( {n\over {1 \ {\rm cm}^{-3}} } \right)
\left( {v\over {40 \ {\rm km \ s}^{-1}} } \right)^{-3} {\rm erg \ s}^{-1}
\end{eqnarray}
\noindent where $\eta$ is the relativistic efficiency ($\eta \simeq 0.18$
for a neutron star radius of $r_* \simeq 12$ km and a mass of $M_* \simeq
1.4\, M_\odot$) and $c$ is the light velocity.
If the star moves subsonically relative to the ISM, equation (1) remains
valid provided that the ISM sound speed $v_s$ is used in place of $v$
($v_s\simeq 10\sqrt{(T/10^4 \, {\rm K})} \ {\rm km \, s^{-1}}$).

As shown by equation (1), in order to investigate the detectability
of ONSs, it is crucial to have knowledge of the density distribution of
the ISM and the velocity distribution of ONSs in the Galaxy.
Calculations of the temporal evolution of the ONS distribution function
have been carried out by many authors (Paczy\'nski 1990; Hartmann,
Epstein \& Woosley 1990; Blaes \& Rajagopal 1991; Blaes \& Madau 1993;
Zane et al. 1995a).
Since we are interested to investigate the detectability of
ONSs in the solar neighbourhood and close to the Galactic plane,
in the following we will use the analytic approximation to the
ONS distribution computed by Zane et al. (1995a).

The structure of the Local Interstellar Medium has been widely
investigated: within $\sim$ 50--100 pc from the Sun, the gas is tenuous
($n \simeq 0.05-0.07$ cm$^{-3}$) and warm ($T \simeq 10^4$ K), although
the presence of a number of relatively high density regions ($n \sim 1$
cm$^{-3}$) can rise the observed column density up to $\sim 2 \times
10^{20}$ cm$^{-2}$ along certain line of sights, such as those in the
direction of Cygnus Rift and Cygnus OB7 (Frisch and York 1983; Paresce
1984;  Welsh et al. 1994; Diamond et al. 1995).
On much larger scales, according to Dickey \& Lockman (1990), the average
density in the Galactic plane is $n \simeq 1$ cm$^{-3}$ with a scale
height variable with the distance from the Sun, although there are a
number of tenuous ($n \simeq 0.01$ cm$^{-3}$) hot ($T \simeq
10^5$--$10^6$ K) bubbles.
Since Cygnus Rift and Cygnus OB7 are very close to the Galactic plane
($-5\degr < b < 5\degr$), in the following the ISM will be approximately
described as a piecewise constant density medium with $n_H = 0.645$
cm$^{-3}$ for $d < 100$ pc and $n_H = 1$ cm$^{-3}$ elsewhere. For the
sources embedded in the clouds and the background sources, the
contribution of the cloud material to the total absorption will be added.

A thorough investigation of the emitted spectrum turns out to be a further
key ingredient in order to correctly address the issue of
detecting ONSs, since the choice of the energy bands where to
look for and the absorption of the interstellar medium are strongly
related to the ONSs emission properties.
As shown by Alme \& Wilson (1973), if
binary Coulomb collisions between the infalling ions and the atmospheric
electrons dominate, as it is expected at very low accretion rates, the
accretion flow can be stopped at several Thomson depths and the resulting
spectrum can be thermalized at a temperature approximately equal to the
star effective temperature
\begin{eqnarray}
T_{eff} & = & \left( {L\over {4\pi r_*^2 f_s \sigma}} \right)^{1/4}
\simeq 90 \left( {f_s\over {10^{-3}}} \right)^{-1/4}\nonumber \\
& &\left( {{r_*}\over {10^6 \ {\rm cm}}} \right)^{-1/2}
\left( {M_*\over {M_\odot}} \right)^{1/2}\nonumber \\
& & \left( {n\over {1 \ {\rm cm}^{-3}} } \right)^{1/4}
\left( {v\over {40 \ {\rm km \ s}^{-1}} } \right)^{-3/4} \ {\rm eV}
\end{eqnarray}
\noindent where $f_s$ is the fraction of the surface area which undergoes
accretion (see below).
Then, ONSs accreting directly from the ISM emit typically in the
ultraviolet and soft X--ray bands. The low bolometric luminosity ($L \sim
10^{30}-10^{31}$ erg s$^{-1}$) and the softness of the spectrum ($T_{eff}
\sim 90$ eV) explain the difficulty in observing an isolated ONS.
Previous investigations were restricted to
the calculation of the emitted spectrum at high luminosities. A detailed
numerical analysis of the spectral properties of unmagnetized neutron stars
accreting well below the Eddington limit has been recently presented by
Zampieri et al.\/ (1995). The emergent spectrum turns out to be
significantly hardened with respect to a black body at the star effective
temperature, with a broad maximum shifted toward higher frequencies
by a factor $\sim$ 3 at $L \sim 10^{30}$ erg s$^{-1}$.
This fact is due entirely to the frequency dependence
of the free--free opacity, for which higher energy photons decouple
at larger depths and temperatures in the neutron star atmosphere.

In the following, we will consider two possibilities for the emitted flux:
first, black body emission at the neutron star effective temperature
(see equation 2)
which has been frequently used in the previous investigations and can be
regarded as a useful approximation;
second, the spectra computed by Zampieri et al. (1995).
In addition, we assume that a relic magnetic field is present and that
the accretion flow is channeled into the polar caps. If we neglect
all the radiative effects produced by the magnetic field (on this regard
see e.g. Miller 1992; Shibanov et al. 1992; Nelson et al.
1995), the main consequence is to limit the size of the emitting region
by a factor $f_s = A_c/2\pi r_*^2$, where $A_c = \pi r_*^3/r_A$ is the
area of the polar cap and $r_A$ is the Alfv\'en radius. For $B \simeq
10^9$ G, as we consider here, $f_s \simeq 10^{-3}$. This fact will
produce a hardening of the spectrum with respect to the unmagnetized case
with the same luminosity, since the flux emitted per unit surface is
$F_{mag} = L/2A_c > F_{unmag} = L/4\pi r_*^2$.

\begin{table*}
\caption{Expected number of ONSs detectable in the direction of
Cygnus Rift as a function of threshold ($N_{ONS}^{tot} = 10^9$, $B = 10^9$ G).
Values in the second column (hardness ratio $HR$) and
in the last two columns refer to the Zampieri's spectrum and to a
black body (values in brackets). $d_j \div d_{j+1}$ is the interval
of distance considered, $(N_{ONS}^{theor})_{[d_j,d_{j+1}]} $
the expected number of ONSs above threshold $S$ detectable in the
direction of the whole cloud area (see the Appendix) and 
$N(>S)_{[d_j,d_{j+1}]}$ the expected number of sources in the fraction
of the cloud $f$ covered by ROSAT pointings (see equation (3)).}
\begin{tabular}{lccccc}
threshold $S$ & coverage $f$ & hardness ratio $HR$ & $d_j \div d_{j+1}$ &
$(N_{ONS}^{theor})_{[d_j,d_{j+1}]}$
& $N(>S)_{[d_j,d_{j+1}]}$ \\
(c s$^{-1}$) & (\%) & & (pc) & & \\

  0.001 & 0.63 & & & & \\

 & &  --0.86$\div$0.85 (--0.98$\div$0.69) &  30$\div$100 & 3 (3) & 0 (0) \\

 & &  0.06$\div$0.94  (--0.85$\div$0.90) &  100$\div$300 & 64 (43) & 1 (1) \\

 & &  0.82$\div$0.97  (0.45$\div$0.95)	&  300$\div$662 &  270 (113) &	7 (2) \\

 & &  0.94$\div$0.99 (0.87$\div$0.99)   &  662$\div$738 [Rift] &  297 (215)
&  11 (6) \\

 & &  0.99 (0.98$\div$0.99)   &  738$\div$2500 (738$\div$2080) & 360 (127)
&  5 (2) \\

  0.002	 & 2.52 & & & & \\

 & &  --0.81$\div$0.85 (--0.98$\div$0.69) &  30$\div$100 &  3 (3)  & 0 (0) \\

 & &  0.07$\div$0.94  (--0.83$\div$0.90) &  100$\div$300 &  52 (33) & 1 (1) \\

 & &  0.84$\div$0.97  (0.53$\div$0.95) &  300$\div$662 &  167 (73) & 6 (2) \\

 & &  0.95$\div$0.99 (0.89$\div$0.99) &  662$\div$738 [Rift]   &  250
(170) &  10 (6) \\

 & &  0.99 (0.98$\div$0.99) &  738$\div$2000 (738$\div$1630)  &	152 (52)  &  
4 (1) \\

   0.004  & 3.15 & & & & \\

 & &   --0.74$\div$0.85 (--0.97$\div$0.69) &  30$\div$100 & 3 (3) & 0 (0) \\

 & &   0.17$\div$0.94 (--0.79$\div$0.90) &  100$\div$300 & 39 (24) & 1 (1) \\

 & &   0.86$\div$0.97 (0.62$\div$0.95)	&  300$\div$662 & 96 (46) & 4 (1) \\

 & &   0.95$\div$0.99 (0.90$\div$0.99) &  662$\div$738 [Rift] &  191 (124)
&  9 (5) \\

 & &   0.99 (0.98$\div$0.99) &  738$\div$1500 (738$\div$1260) & 47 (17) & 1
(1) \\

  0.008	 & 4.41 & & & & \\

& &  --0.64$\div$0.85 (--0.96$\div$0.69) &  30$\div$100 & 3 (3) & 0 (0) \\

 & &  0.33$\div$0.94 (--0.68$\div$0.90) &  100$\div$300 &  26 (16) & 1 (1) \\

 & &  0.88$\div$0.97 (0.69$\div$0.95)  &  300$\div$662 &  51 (27) & 2 (1) \\

 & &  0.96$\div$0.99 (0.92$\div$0.99) & 662$\div$738 [Rift] & 129 (85) & 7
(4) \\

 & &  0.99 (0.98$\div$0.99) & 738$\div$1100 (738$\div$960) &  12 (4) & 1
(0) \\

  0.01	& 5.04 & & & & \\

 & & --0.60$\div$0.85 (--0.96$\div$0.69) & 30$\div$100  & 2 (2) & 0 (0) \\

 & & 0.35$\div$0.94 (--0.64$\div$0.90) & 100$\div$300  & 23 (14) & 1 (1) \\

 & & 0.89$\div$0.97 (0.71$\div$0.95)  & 300$\div$662  & 42 (23) & 2 (1) \\

 & & 0.96$\div$0.99 (0.92$\div$0.99) &  662$\div$738  [Rift] & 110 (74)  & 6
(4) \\

 & & 0.99 (0.98$\div$0.99) &  738$\div$1030 (738$\div$880) &  8 (2) &  0
(0) \\

\end{tabular}
\end{table*}

\begin{table*}
\caption{Expected number of ONSs detectable in the direction of
Cygnus OB7 as a function of threshold ($N_{ONS}^{tot} = 10^9$, $B = 10^9$ G).
Values in the second column (hardness ratio $HR$) and
in the last two columns refer to the Zampieri's spectrum and to a
black body (values in brackets). $d_j \div d_{j+1}$ is the interval
of distance considered, $(N_{ONS}^{theor})_{[d_j,d_{j+1}]} $
the expected number of ONSs above threshold $S$ detectable in the
direction of the whole cloud area (see the Appendix) and 
$N(>S)_{[d_j,d_{j+1}]}$ the expected number of sources in the fraction
of the cloud $f$ covered by ROSAT pointings (see equation (3)).}
\begin{tabular}{llcccc}
threshold $S$ & coverage $f$ & hardness ratio $HR$ & $d_j \div d_{j+1}$ &
$(N_{ONS}^{theor})_{[d_j,d_{j+1}]}$
& $N(>S)_{[d_j,d_{j+1}]}$ \\

  0.001 & 1.55 & & & & \\

  & &  --0.86$\div$0.85 (--0.98$\div$0.69) & 30$\div$100 &  3 (3) & 0 (0) \\

  & &  0.06$\div$0.94 (--0.85$\div$0.90) & 100$\div$300 &  64 (43) &  2 (1) \\

  & &  0.82$\div$0.99 (0.45$\div$0.99)	& 300$\div$739 &  136 (56) & 4 (1) \\

  & &  0.95$\div$0.99 (0.89$\div$0.99) &  739$\div$861 [OB7] & 216 (140)
& 6 (4) \\

  & &  0.99 (0.99) &  861$\div$2100 (861$\div$1700) &  67 (22) & 1 (0) \\

  0.002 & 2.32 & & & & \\

& &  --0.81$\div$0.85 (--0.98$\div$0.69) &  30$\div$100	 &  3 (3) & 0 (0) \\

  & &	0.07$\div$0.94 (--0.83$\div$0.90) &  100$\div$300 &  52 (33) & 2 (1) \\

  & &	0.84$\div$0.99 (0.53$\div$0.99)	 &  300$\div$739 &  82 (36) & 3 (1) \\

  & &	0.96$\div$0.99 (0.90$\div$0.99) &  739$\div$861 [OB7] &  167 (105) &
6 (3) \\

  & &	0.99 (0.99) &  861$\div$1600 (861$\div$ 1300) &	22 (7) & 1 (0) \\

  0.004 & 3.87 & & & & \\

  & &  --0.74$\div$0.85 (--0.97$\div$0.69) &  30$\div$100 &  3 (3) & 0 (0) \\

  & &	0.17$\div$0.94 (--0.79$\div$0.90) & 100$\div$300 &  39 (24) & 2 (1) \\

  & &	0.86$\div$0.99 (0.62$\div$0.99)	 & 300$\div$739 &  46 (23) & 2 (1) \\

  & &	0.96$\div$0.99 (0.92$\div$0.99) &  739$\div$861 [OB7] &  117 (73)
& 5 (3) \\

  & &	0.99 (0.99) &  861$\div$1200 (861$\div$ 990) &	5 (1) & 0 (0) \\

  0.008 &  3.87 & & & & \\

  & &	--0.64$\div$0.85 (--0.96$\div$0.69)  &  30$\div$100  &  3 (3) & 0 (0) \\

  & &	0.33$\div$0.94 (--0.68$\div$0.90)  & 100$\div$300  &  26 (16) & 1 (1) \\

  & &	0.88$\div$0.99 (0.69$\div$0.99)	  & 300$\div$739  &  25 (13) & 1 (1) \\

  & &	0.96$\div$0.99 (0.93$\div$0.99) &  739$\div$861 (739$\div$ 839) [OB7]
& 72 (45) & 3 (2) \\

  & &	0.99  &	 861$\div$914 &	 0 \\

  0.01 &  4.64 & & & & \\

  & &	--0.60$\div$0.85 (--0.96$\div$0.69)  &  30$\div$100 &  2 (2) & 0 (0) \\

  & &	0.35$\div$0.94 (--0.64$\div$0.90)  & 100$\div$300 &  23 (14) & 1 (1) \\

  & &	0.89$\div$0.99 (0.71$\div$0.99)	  & 300$\div$739 &  20 (11) & 1 (1) \\

  & &	0.97$\div$0.99 (0.93$\div$0.99) &  739$\div$861 (739$\div$824) [OB7]
& 65 (33) &  3 (2) \\

\end{tabular}
\end{table*}

\subsection{Expected number of ONSs}

The technique which has been used to calculate the expected number of ONSs
in the molecular clouds is the generalization of a similar
procedure introduced by Zane et al. (1995b) and is described in detail in
the Appendix. With this technique it
is possible to calculate also the expected number of foreground and
background ONSs accreting from the average ISM, which are seen in the
direction of the clouds but are not embedded within them. This aspect
deserves particular notice since, as shall see, they are
expected to be quite numerous at low sensitivity limits and could be
distinguished from the ONSs embedded in the molecular clouds only by
means of their observed spectral properties.

The characteristic parameters of the two clouds are quoted in Table 1.
Results of the expected number of sources observable in Cygnus Rift,
Cygnus OB7 and also in other spatial regions in the cloud
directions are presented in Tables 2 and 3 (fifth column) for
different values of the sensitivity limit $S$ ($N_{ONS}^{tot} =
10^9$, $B = 10^9$ G). These numbers are calculated assuming the
spectrum of Zampieri et al. (1995) and a black body spectrum (values
in brackets).
The fraction of clouds which has been covered by the ROSAT pointings at a
given sensitivity depends mainly on the exposure time.
In the second column of Tables 2 and 3 we quote the fractional 
coverage of the cloud areas $f$ as a function of threshold. As can be seen,
the covering increases from $\sim$ 1\% at $10^{-3}$ c s$^{-1}$
to $\sim$ 5\% at $10^{-2}$ c s$^{-1}$. 
To compare the theoretical estimates with the actual number of
unidentified X--ray sources detected in the ROSAT pointings
of our sample, for each interval of distance we have computed the
expected number of detectable sources according to the following
formula
\begin{eqnarray}
   N(>S_i)_{[d_j,d_{j+1}]} & = & 
  (N_{ONS}^{theor})_{[d_j,d_{j+1}]}^i f_i \nonumber \\ & &
+ (N_{ONS}^{theor})_{[d_j,d_{j+1}]}^{i+1} (f_{i+1}-f_i) + \nonumber \\ 
   && + ... +
\nonumber \\ & &
+ (N_{ONS}^{theor})_{[d_j,d_{j+1}]}^{i_{max}} (f_{i_{max}}-f_{i_{max}-1}) 
\, ,
\end{eqnarray}
where $(N_{ONS}^{theor})_{[d_j,d_{j+1}]}^i$ is the expected number of 
ONSs above threshold $S_i$ detectable in the direction of the whole
cloud areas and $f_i$ is the fractional coverage of the pointings
at the same threshold (see Tables 2 and 3).
Here $S_1 = 10^{-3}$ c s$^{-1}$, $S_2 = 2 \times
10^{-3}$ c s$^{-1}$, $S_3 = 4 \times 10^{-3}$ c s$^{-1}$,
$S_4 = 8 \times 10^{-3}$ c s$^{-1}$, $S_{i_{max}} = S_5 = 10^{-2}$ c s$^{-1}$.
The first term on the right hand side of equation (3)
gives the number of sources observable
in the pointings with limiting sensitivity $S_i$, while the other terms
account for the contributions from pointings at lower sensitivity.
Finally, the total expected number of sources $N(>S_i)$ is obtained
integrating $N(>S_i)_{[d_j,d_{j+1}]}$ over distance and
is shown in Figure 2 (summing the contributions from the two clouds).

In the fraction of the clouds actually observed by ROSAT,
we expect the detection of 10--24 sources (the exact value depends on the
assumption on the emitted spectrum)
at the sensitivity limit of $1-2 \times 10^{-3}$ c s$^{-1}$ in the
direction of Cygnus Rift and of 5--13 sources in the direction of Cygnus
OB7. Among these sources, 6--11 sources are expected to be really embedded
in Cygnus Rift and 3--6 in Cygnus OB7. The other objects are foreground
or background sources; in particular, we estimate that 3--8 and 2--6
foreground ONSs should be detectable in the direction of Cygnus Rift and
Cygnus OB7, respectively. We note that at high thresholds $10^{-2}$
c s$^{-1}$ the detection of a significant number of sources would be 
expected.

Then, in agreement with previous investigations (Blaes \& Madau 1993;
Colpi, Campana \& Treves 1993; Zane et al. 1995a)
Cygnus Rift and Cygnus OB7 are expected to be particularly favourable sites
for the observability of ONSs with ROSAT PSPC. In addition, we have found
that in their direction a significant number of foreground and background
ONSs should be observable.

\subsection{Expected emission properties}

In principle we could compare
directly the theoretical spectral distribution with the observed one
but, as we shall see,
the sources selected in our sample have too few photons to extract a
spectrum. Then, in order to compare our theoretical expectations with the
observations, in each range of distances considered we have simulated
PSPC spectra by folding the theoretical model with interstellar absorption,
detector effective area and response matrix, and
calculated the PSPC hardness ratio, defined by
\begin{equation}
HR = { {N(41-240) - N(11-40)}\over
       {N(41-240) + N(11-40)} } \, ,
\end{equation}
where $N(11-40)$ and $N(40-240)$ are the count rates in the PSPC channel
ranges 11--40 and 41--240, corresponding roughly to the energy bands
0.1--0.4 keV and 0.4--2.4 keV.
We have repeated the calculation assuming either the spectrum computed by
Zampieri et al. (1995) or a black body at the neutron star effective
temperature. Results are presented in Tables 2 and 3, where the lowest
and highest values of the hardness ratio is reported for each interval
of distance (the highest value refers to the model with maximum
luminosity). It is interesting to
note that, for distances above $\sim 600$ pc for the
black body and $\sim 300$ pc for
the synthetic spectra, the absorption of the interstellar medium and the
cloud material causes the hardness ratio to be very close to unity.
Then, although ONSs are relatively soft sources, above these distances they
should appear significantly hardened.

We have estimated also the apparent visual magnitude of ONSs below 100 pc
from the Sun and find that $m_V \ga 32$. Then, these sources should lack
of optical counterparts in the digitized red plates of the Palomar All Sky
Survey (POSS, limiting magnitude
$m_r \sim$ 20). According to recent findings by Blaes, Warren \& Madau (1995),
for ONSs embedded within the clouds and considering
polar cap emission the reprocessing of the UV--soft X--ray radiation
by the surrounding gas might increase the emitted flux at optical wavelengths
by 1--2 magnitudes. However, even in this case the very low visual magnitude
prevents any possibility of optical identification for sources beyond 100 pc.
Then, we will use the lack of optical counterparts in the digitized POSS
plates as a distinguishing criterion for selecting ONSs candidates
among ROSAT sources.

Inserting in equation (A2) the appropriate effective areas and frequencies
(taken from the world wide web site of the Center for Extreme Ultraviolet
Astrophysics), we have calculated also the expected count rates
in the Lex band (67--178 \AA) of the Extreme Ultraviolet Explorer Deep Survey
(EUVE DS). Below $\sim$ 250--300 pc, the count rate turns out to be smaller
than $10^{-5}$ c s$^{-1}$ even for the more luminous sources and no
detectable UV flux can be observed at Earth.

Finally, comparing the observed hardness ratios with the values quoted in
Tables 2 and 3 could be used to discriminate ONSs among other
optically unidentified X--ray ROSAT sources.



\section{Data analysis}

\subsection{The sample}

We selected from the ROSAT Public Data archive PSPC pointings in the
direction of the Cyg OB7 and Cyg Rift molecular clouds. A description of
the satellite and the detector can be found in Tr\"umper (1983) and
Pfeffermann et al. (1986). The 28 analyzed areas are shown in Figure 1,
overlaid onto the CO contour maps from Dame \& Thaddeus (1985). The
relevant data on the pointings are reported in Table 4.
A few pointings lie just outside the contour of the clouds in Figure 1,
but have been nevertheless included in the sample.
The total exposure time is 126603 seconds.
Only the central region of the PSPC (radius 20$'$) has been considered
for the analysis, to avoid problems due to the poor knowledge of the
point spread functions outside that region. The areas of the two clouds
covered by ROSAT pointings are 2.1 square degrees (Cyg
OB7) and 5.6 square degrees (Cyg Rift), which correspond to 4.6\% and
5.0\% of the total areas of the clouds, respectively (adopting the values
quoted in Dame \& Thaddeus 1985).
Because of the nature of the pointings, the fractional coverage of the cloud
areas is limited and varies with threshold (see Tables 2 and 3).
Although at the lowest sensitivity limits (1--2 $10^{-3}$ c s$^{-1}$) only
1-2\% has been covered, the theoretical expectations (see again Tables 2 and 3)
indicate that this could be sufficient for detecting ONS candidates.
Moreover, from the systematic analysis of these data
a search methodology can be outlined.

\begin{table}
\caption[o]{Observation log. The first two columns are the identifier
of the pointing in this paper and the ROSAT observation number.
Galactic coordinates are in decimal degrees. N$_X$ is the number of
detected X--ray sources and N$_{GSC}$ is the number of GSC stars used
for the boresight correction (noB indicates one star or less, see text).}
{\small
\label{tab1}
\begin{center}
\begin{tabular}{lcccc|cc}
Label & ROR \# & Exp. (s) & l$_{II}$ & b$_{II}$ &
	     N$_X$ & N$_{GSC}$ \\
\hline\hline
\multicolumn{5}{c}{Cyg OB7}\\
A  &	200427&	  382.0	 &    91.4  &	  1.1 &	 1 & noB \\
K  &	500143&	 8260.0	 &    93.3  &	  6.9 &	 9 &  4	 \\
L  &	500254&	 4297.0	 &    93.7  &	--0.3 &	 1 & noB \\
S  &	300033&	 4331.0	 &    89.8  &	--0.1 &	 5 & noB \\
T  & 300033\_1&	 2854.0	 &    89.8  &	--0.1 &	 6 & 3	 \\
W  &	500195&	 9137.0	 &    94.0  &	  1.0 &	 7 & 5	 \\
b  &	900194&	 4166.0	 &    89.4  &	--0.7 &	 3 & 2	 \\
\hline
\multicolumn{5}{c}{Cyg Rift}\\
B  &	200057&	 1846.0	 &    76.6  &	  1.4 &	 4 &  3	 \\
C  &	200058&	 1931.0	 &    76.6  &	  1.4 &	 2 &  2	 \\
D  &	200062& 14030.0	 &    76.6  &	  1.4 & 17 & 14	 \\
E  &	200063&	 2073.0	 &    76.6  &	  1.4 &	 3 &  3	 \\
F  &	200109&	 3587.0	 &    80.2  &	  0.8 & 11 &  9	 \\
G  &	200114&	 3508.0	 &    85.7  &	  1.5 &	 1 & noB \\
H  &	200564&	 8950.0	 &    74.3  &	  1.1 &	 5 &  3	 \\
I  &	500084&	 2470.0	 &    69.7  &	  1.0 &	 2 &  2	 \\
J  &	500085&	 2879.0	 &    83.0  &	--0.3 &	 2 &  2	 \\
M  &	900157&	 4054.0	 &    78.2  &	  2.0 &	 2 &  1	 \\
N  &	900158&	 4123.0	 &    76.0  &	  0.4 &	 1 & noB \\
O  &	200154&	 1261.0	 &    75.8  &	  1.3 &	 0 & --	 \\
P  &	200412&	 1265.0	 &    78.2  &	  0.1 &	 1 & noB \\
Q  & 200412\_1&	 2228.0	 &    78.2  &	  0.1 &	 2 & noB \\
R  &	200702&	 4409.0	 &    85.7  &	--0.3 &	 7 &  5	 \\
U  &	400157&	 2996.0	 &    77.0  &	  3.2 &	 3 &  2	 \\
V  &	500124&	 5979.0	 &    70.7  &	  1.2 &	 1 & noB \\
X  &	500203&	 3641.0	 &    84.2  &	--0.8 &	 1 & noB \\
Y  &	500206& 10323.0	 &    78.4  &	  2.8 & 13 &  6	 \\
Z  &	500207&	 9319.0	 &    77.9  &	  3.5 &	 8 &  5	 \\
a  &	500208&	 2304.0	 &    77.5  &	  4.0 &	 2 & noB \\
\hline
Total	&     & 126603	 &	    &	      & 120 &	\\
\end{tabular}
\end{center}   }
\end{table}

   \begin{figure*}
\vskip 8 true cm
      \caption{Map of the analyzed ROSAT PSPC pointings. Each circle is
      20$'$ radius. The black circles correspond to pointings containing
      ONS candidates (see text). The contours of the clouds are from Dame
      \& Thaddeus (1985).}
   \end{figure*}

\subsection{The analysis procedure}

Each pointing has been extracted from the ROSAT Public Archive at MPE
Garching and analyzed separately by running a set of semi-automatic
programs. The standard detection technique in the EXSAS package
(Zimmermann et al. 1994) has been applied. First a background map is
produced by removing all possible sources (identified by means of a
sliding window technique) and running a two dimensional spline fit to the
data. Then a Maximum Likelihood (ML) algorithm is applied (see Cruddace,
Hasinger \& Schmitt 1988) to detect significant deviations from the
estimated background distribution. The threshold for detection has been
set to the conventional value of ML=10, which corresponds to a chance
detection probability of 4.5$\times 10^{-5}$ for a single trial.
This technique is repeated for the data in three different PHA channel
ranges: total (T: channels 11--240, 0.1--2.4 keV), soft (S: channels
11--40, 0.1--0.4 keV) and hard (H: channels 41--240, 0.4--2.4 keV).
In case of a detection in more than one band, only the band with the
highest value of ML has been considered.

It is known that the ML technique becomes problematic in presence of
large regions of extended emission, as it is likely to find in the
pointing directions considered here, lying on the galactic plane.
Therefore, all the detected sources have been visually inspected in the
X--ray images and obvious spurious detections have been discarded.

   \begin{figure}[htbp]
     \vskip 7.7 truecm
      \caption{The $\log N$--$\log S$ distribution for all the
	       sources detected by ROSAT (continuous line) and all the
	       optically unidentified sources detected by ROSAT
	       (dashed line). Also shown is the expected number of 
               ONSs, $N(>S)$, in the fraction of the clouds covered by ROSAT
               pointings (dash--dotted line; Zampieri's spectrum).
	      }
	 \label{Fig2}
   \end{figure}


In the cases where two or more pointings were coaligned, they were analyzed
separately: the sources have
been cross-correlated and in case of coincidences only the source with
the highest detection ML has been considered. Eleven sources have been
eliminated in this way, bringing the total number of sources to 109.
The $\log N$--$\log S$ distribution for the sources detected by ROSAT
is shown in Figure 2. There are a few
strong sources, but most of them are below 0.01 c s$^{-1}$. Note
that the targets of the pointings have not been removed.

\subsection{Optical identification}


The positions of the detected X--ray sources had to be corrected for an
offset due to a residual systematic uncertainty in the position
determination (the so-called boresight correction).
In order to do this, preliminary identifications have been performed
using the HST Guide Star Catalog (GSC; Lasker et al. 1990).  The X--ray
sources and GSC stars have been cross-correlated and the best shifts to
the X--ray positions have been determined by means of a Maximum
Likelihood technique. The shift so obtained has been considered only if
at least two sources were identified in this way (with the exception of
pointing S, where one of the two sources identified with GSC entries had
multiple bright optical counterparts). For the remaining pointings, a
6$''$ (1$\sigma$) error has been quadratically added to the error radii
of the detected sources (K\"urster \& Hasinger 1992). In Table 4 a
summary of the X--ray detections and boresight corrections is reported.
An X--ray source is considered to be identified if a GSC star falls into
its 90\% positional error circle. In case of more than one optical
identification, the brightest counterpart has been considered.

67 X--ray sources were identified with entries in the GSC, which
has a loosely defined magnitude limit of $m_V \sim 14$. For the
remaining 42 sources, we used the Digitized
Sky Survey (DSS; Postman et al. 1995), which consists of the POSS red
plates. The magnitudes of the brightest stars in the error boxes have
been estimated by using the empirical magnitude--diameter relation of
King \& Raff (1977) appropriate for red POSS plates. The values so
obtained have a large uncertainty (up to one magnitude), but it is
sufficient for our purpose. Only in three cases the possible optical
counterpart was not bright enough to allow an estimate of the magnitude,
the image being not saturated. In Figure 3 we show the distribution of
magnitudes of the optical counterparts for the two samples, GSC and DSS.
As one can observe there is a considerable number of bright stars, since
there are bright OB associations in the direction of most of the
pointings; indeed a large number of the counterparts for which we were able
to obtain a spectral type were of types O and B.
The presence of strong optical candidates for many X--ray
sources ensures the robustness of the boresight process.

After the optical screening, 7 sources
were left for which no optical counterpart could be seen in the POSS
red plates, to a limiting magnitude $m_r\sim$ 20.
Two of these were coincident with the original targets of the
pointings and were therefore excluded. The brightest of the remaining
sources was subsequently identified with Cyg X--3, so
4 final candidates remained.

\begin{table*}
\caption{ONS candidates. The columns are: source name,
pointing identifier, celestial coordinates, 90\% error radius, Maximum
Likelihood of existence, PSPC total count rate (channels 11--240) and
hardness ratio (defined as $HR =$ (H--S)/(H+S), where S and H
are the counts in the channel intervals 11--40 and 41--240 respectively.
}
\begin{flushleft}
\begin{tabular}{rcllcccc}

Name	& P. & ~~$\alpha$ (2000) & ~~$\delta$ (2000) & $\Delta r$($''$) & ML &
	 c ksec$^{-1}$  & {\it HR} \\
\hline\hline

%
%
%
%
Rift--1 & D  & 20\hh 19\mm 27.96\ss & +38\degr 38\arcmin 27.6\arcsec
	     & 10.5	  & 24.7  &  2.7$\pm$0.6  & 0.29$\pm$0.23 \\

Rift--2 & Y  & 20\hh 19\mm 47.05\ss  & +41\degr 12\arcmin 01.3\arcsec
	     & 19.2	  & 10.9 &   3.2$\pm$0.9  & $>$0.28	  \\

OB7--1	& K  & 20\hh 53\mm 07.35\ss  & +55\degr 13\arcmin 29.0\arcsec
	     & 13.4	  & 17.7 &   1.5$\pm$0.5  & $>$0.44	  \\

OB7--2	& W  & 21\hh 25\mm 31.12\ss  & +51\degr 48\arcmin 31.5\arcsec
	     &	9.2	  & 35.2 &   2.4$\pm$0.6  & $>$0.84	  \\

\end{tabular}
\end{flushleft}
\end{table*}

\subsection{ONS candidates}

For the four sources with no optical counterpart,
PSPC count rates in the different bands have been
extracted manually from the data, with an accurate estimate of the
background, in some case contaminated by extended emission. The relevant
information about the candidates (two in each cloud) can be found in
Table 5. As it can be seen from Table 4, all four sources come from
pointings with long exposure times and with a large number of GSC
identifications, which ensures an accurate boresight correction.
Source OB7--2 lies just outside the outer rim of the cloud (see Figure
1), but we decided not to discard it from the sample.

   \begin{figure}[htbp]
     \vskip 7.7 truecm
      \caption{Distribution of optical magnitudes for the possible
	       counterparts of the X--ray sources. GSC magnitudes are in
	       the $V$ band, POSS magnitudes in the red (see text).
	       The lack of POSS
	       stars at magnitude 19 is a probable artifact of the
	       procedure for magnitude estimate.}
      \label{Fig3}
    \end{figure}

Source Rift--2 might have counterparts at different wavelengths. The POSS
plate shows a very faint ring--like structure ($\sim 20''$ diameter) in
the X--ray error box, reminiscent of a supernova remnant or a planetary
nebula. The position of the source is compatible with that of the IRAS
source 20179+4102, which has a 100 $\mu$m flux of 75 Jy (IRAS Catalogs
and Atlases 1985). Putting the distance at 0.7 kpc,
this translates roughly to a luminosity of $\sim 10^{35}$ erg s$^{-1}$.
At the same distance the radius of the optical ring would be $\sim 0.1$
pc. If the ring and the IRAS sources were the same object, it would be
too small and faint to be a supernova remnant, while the planetary nebula
hypothesis could not be ruled out. In absence of a clear identification
we do not exclude the source from our candidates list.


\section{Discussion}

As can be seen from Tables 2, 3 and 5, our
predicted number of sources is larger than the number of unidentified
objects detected by ROSAT. In particular, Figure 2 shows
the expected number of ONSs and the observed number of unidentified
objects in our sample as a function of threshold. By comparing the two
curves, it is clearly visible that predictions are too ``optimistic'',
although it should be stressed that, by purpose, 
theoretical calculations have usually been performed under the most
favourable assumptions.
The fact that the observed number of ONS candidates is 
smaller by a factor 3--10 indicates 
that the range of parameters of our model needs to be
more constrained. As already discussed, there are several sources
of uncertainties which may influence theoretical predictions, such as
the velocity distribution and density of ONSs, the strength of the
relic magnetic field (related to the possible inhibition of
accretion), the neutron star birth rate, the
total number of ONSs, the indetermination in the
exact value of the gravitational capture radius $r_{gc}$ and the 
validity of the fluid approximation for the average ISM.
Scattering by molecular clouds and spiral arms (dynamical heating; Wielen
1977) may boost a fraction of low velocity stars to higher speeds,
reducing the number counts up to an order of magnitude (Madau \& Blaes
1994). Also, evidence has been found (Lyne \& Lorimer 1994) for a
neutron star mean birth velocity much higher than the one obtained by
Narayan \& Ostriker (1990). Recently, Blaes, Warren \& Madau (1995)
have shown that preheating of the ambient gas to temperatures higher
than $10^4$ K may significantly reduce the accretion rate, hence
decreasing the number of detectable objects.


Our ONSs searching strategy was aimed to select X--ray objects in
molecular clouds which, within their error boxes, do not have any optical
counterpart in the POSS plates. Additional information is clearly needed
to assess the true nature of these objects. The detection of an X--ray
pulsation in the range of 1--1000 s will clearly indicate that we are
observing an ONS; however high magnetic fields are needed
which could prevent accretion and could harden
substantially the emitted spectrum (see e.g. Nelson et al. 1995),
moving the peak emissivity outside the ROSAT energy range.
In absence of X--ray pulsations, a spectral criterion could be used to
individuate ONSs: in the case of low magnetized or unmagnetized ONSs we
expect an X--ray spectrum which show a significant hardening with respect
to a black body at the neutron star effective temperature, as discussed
in Zampieri et al. (1995). An additional feature could be a high column
density, testifying that the X--ray object lies within (or beyond) the
molecular cloud. As can be seen from Tables 2 and 3, the expected hardness
ratios for ONSs embedded in the clouds or beyond them are near to
unity. In this respect, the observed $HR$ of three of our candidates
are consistent with the hypothesis that the sources are located within
the clouds, whereas one source (Rift--1)
cannot certainly lie within the cloud because of its measured value
of the hardness ratio $HR = 0.29 \pm 0.23$ (see Table 5). However,
it should be noted that its spectral properties are consistent with
those of a foreground ONS located at
$\ga$ 100--300 pc (see also Table 2a, $S = 2 \times 10^{-3}$ c s$^{-1}$).
We checked that, at this distances, Rift--1 should not be detectable
even at the lowest sensitivity limits ($\ga 10^{-3}$ c s$^{-1}$)
of EUVE DS.



From the lack of optical counterparts in the digitized POSS plates,
we can estimate a lower limit for the observed
X--ray to optical flux ratio ($F_X/F_V$) for our ONS candidates.
The limiting magnitude in the regions of the four candidates
is $m_r\sim$ 20; using the conversion to magnitude $m_V$ by King and Raff
(1977)
we can adopt this limit as a conservative limit on $m_V$, since only blue
objects would break this assumption and the bluest unabsorbed stars have
B--V=$-0.3$.
For the conversion between ROSAT countrate and flux we adopted the relation
by Fleming et al. (1995), appropriate for sources of stellar nature.
The resulting values for $\log F_X/F_V$ are between --0.4 and +0.1.

Stocke et al. (1995) use the value of $F_X/F_V$
as a chief criterion for determining the nature of the X--ray emitter.
In addition to compact objects, high $F_X/F_V$ 
are associated also with BL Lacs, cluster of galaxies, young supernova
remnant and heavily extincted pre--main sequence stars.
While an extragalactic origin seems unlikely for sources on the galactic
plane in the direction of molecular clouds, the other possibilities still
cannot be excluded.

Follow--up deep observations will be necessary to further investigate
the nature of these ONS candidates. In particular, deep X--ray pointings
in the direction of these four sources would be very useful to measure
with more precision the X--ray spectral properties. Also, UV observations
in the LeX band of EUVE at the lowest sensitivity limits would be
desirable to verify the absence of ultraviolet emission. Finally, deep
optical observations are certainly needed to search for the presence of
possible fainter counterparts within their error circle and to put a
better constraint on the $F_X/F_V$ ratio.

\appendix
\section{Appendix: expected number of ONSs}

In the calculation of the expected number of ONSs
we have used the analytic approximation to the
ONS distribution computed by Zane et al. (1995a), which is based on the
evolution of the F population (characterized by the lowest value of the
mean velocity) of model b of Narayan \& Ostriker (1990) and on the
assumption of a uniform spatial distribution of neutron stars:
\begin{eqnarray}
{{dN_{ONS}}\over {dVdv}} & = & n_{ONS}\, {{dG(v)}\over {dv}} \nonumber\\
& = &
n_{ONS}\, {{d}\over {dv}} \,\Bigl[ {{v/69{\rm \, km \,s^{-1}})^{3.3}}\over{
1+(v/69 {\rm \, km \,s^{-1}})^{3.3}}} \Bigr]\, ,
\end{eqnarray}
where $dV$ is the spatial volume integration element,
$n_{ONS} = 3 \times 10^{-4}\, (N_{ONS}^{tot}/10^9)$ pc$^{-3}$ is the
number density of ONSs averaged in the local region around the Sun
($r \leq 2$ kpc, $|z| \leq 200$ pc), $N_{ONS}^{tot}$ is the
present total number of ONSs in the Galaxy and $G(v)$
is an analytic fit to the cumulative velocity distribution derived by
Zane et al. (1995a).

The count rate (measured at Earth) in a certain spectral interval
($\nu_1$, $\nu_2$) of a star which emits a
monochromatic flux of radiation $F_\nu$ is
\begin{equation}
\widetilde{N}(\nu_1,\nu_2) = \int_{\nu_1}^{\nu_2} f_s {F_\nu\over {h\nu}}
\left( {r_*\over d} \right)^2 e^{-\sigma_\nu N_H} A_\nu d\nu
\end{equation}
\noindent where $d$ is the distance of the source, $\sigma_\nu$ is the
absorption cross--section of the ISM (Morrison \& McCammon 1983) and
$N_H$ the hydrogen column density.
$A_\nu$ is the effective area and $\Delta \nu = \nu_2 - \nu_1$
the bandpass of the detector ($\nu_1 = 0.1$ keV and $\nu_2 = 2.4$ keV
for ROSAT PSPC).

Clearly, at a certain distance $d_j$ an ONS will be detectable if its count
rate $\widetilde{N}$ is above the sensitivity limit $S$ of the detector.
Then, at $d_j$ there exists a limiting value of the star luminosity,
$(L_{min})_{d_j}$
which depends on the emission properties and the absorption of the ISM,
below which an accreting neutron star does not give rise to a
count rate above the threshold of the detector and hence is not observable.
From equation (1) this translates directly into an upper limit for the
star velocity, $(v_{max})_{d_j}$.

Alternatively, a star with a given luminosity $L$, or velocity $v$, will be
observable up to a maximum distance $d_{max}(v)$ at which $\widetilde{N}$
goes below $S$.
Then, the total number of ONSs which can be observed in a certain
interval of distance [$d_j$, $d_{j+1}$] and within a solid angle
$\Sigma$ can be calculated from equation (A1)
summing up all the neutron stars that are contained within the volume
$V(v) = \Sigma [d^3_{max}(v) - d_j^3]/3$ and integrating over $v$:
\begin{eqnarray}
(N_{ONS}^{theor})_{[d_j,d_{j+1}]} & = & n_{ONS}\, \int_{v_{j+1}}^{v_j}
V(v) {{dG}\over {dv}} dv + \nonumber\\
& & + n_{ONS} V(v_{j+1}) G(v_{j+1})
\end{eqnarray}
In equation (A3) the integral goes from the maximum velocity of
detectability at distance $d_{j+1}$, $v_{j+1} = (v_{max})_{d_{j+1}}$, to the
maximum velocity at $d_j$, $v_j = (v_{max})_{d_j}$, whereas
the second term on the right hand side accounts for all the
stars with luminosity above threshold throughout all of the spatial
volume considered. The integral has been evaluated numerically using a
Lobatto quadrature and $d_{max}(v)$ has been interpolated at the
appropriate value of $v$ from a table of entries previously
calculated (an extensive use has been done of a local cubic interpolation
procedure). We note that for each threshold there exists an
absolute upper limit to the distance of detectability which corresponds to
the star with the minimum accretion velocity $v = v_s$ and
maximum emitted luminosity. We note also that in each
interval of distance [$d_j$, $d_{j+1}$] the ISM density must be
constant in order to ensure the regularity of the function $d_{max}(v)$
and hence the correct computation of the integral.

We have estimated the spatial boundaries $r_{in} = r_c - \Delta r_c$ and
$r_{out} = r_c + \Delta r_c$ of Cygnus Rift and Cygnus OB7 ($r_c$ is the
position of the cloud center) by inverting for the cloud width $\Delta
r_c$ the expression $V_c = \Sigma_c \,(r_{out}^3 - r_{in}^3)/3$,
where $V_c$ is the cloud volume and $\Sigma_c$ its apparent
angular surface (taken from Dame et al. 1987; see also Table 1).
Using these boundaries, the expected number of ONSs in each cloud has
been evaluated from equation (A3). Moreover, with the same technique, it
is possible to calculate also the expected number of foreground and
background ONSs accreting from the average ISM, which are seen in the
direction of the clouds but are not embedded within them.

\begin{acknowledgements}

We would like to thank R. Waters and A. Treves for useful discussions.
TB is supported by NWO under grant PGS 78--277; SC gratefully acknowledges
the receipt of an Italian Space Agency fellowship. This research was
supported in part by NSF grant AST 93-15133 and NASA grant 5-2925 at the
University of Illinois.

\end{acknowledgements}

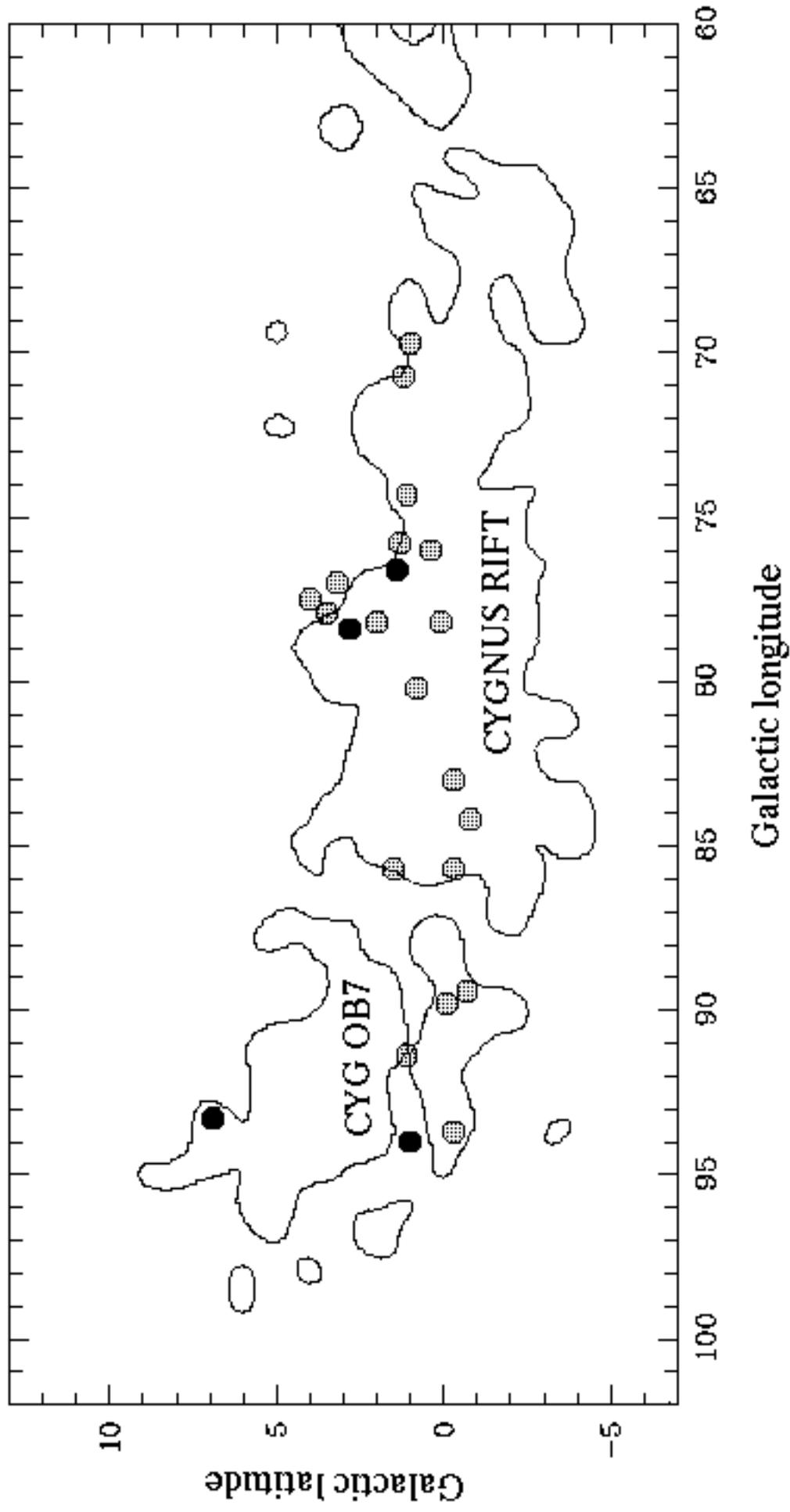
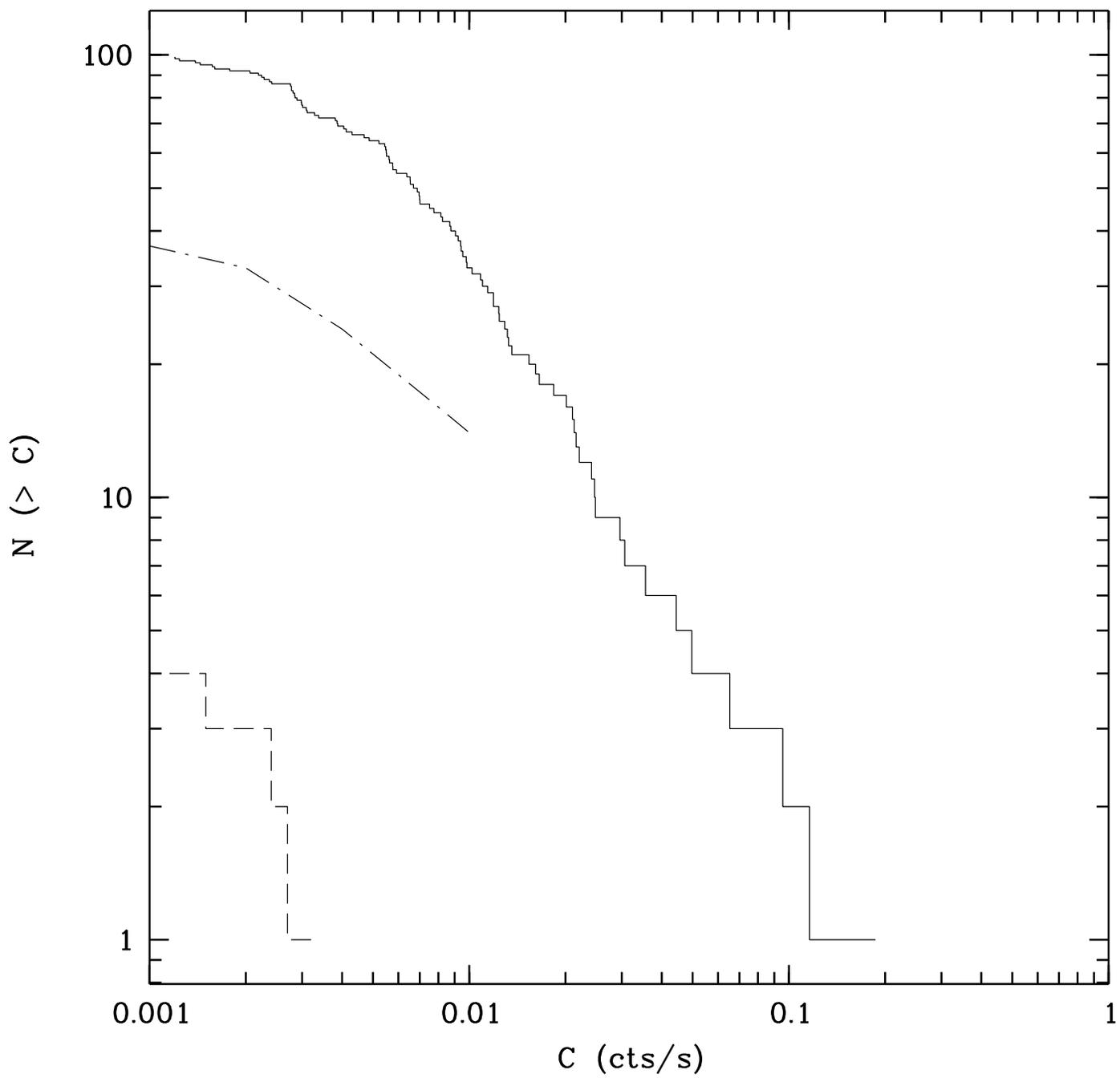
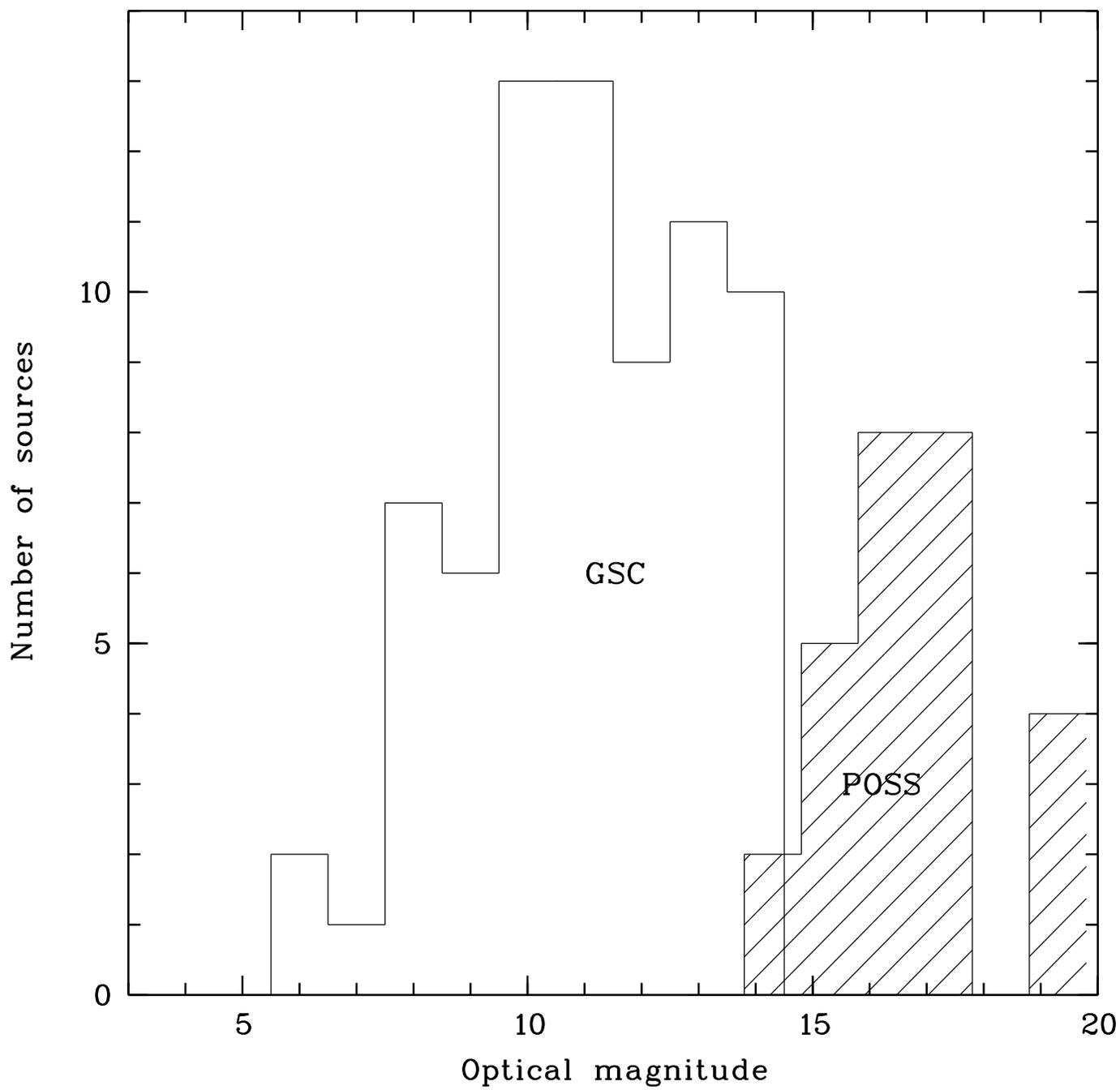
\end{document}